



%
 \hoffset=4pt
 \vsize=9.8in
 \voffset=-0.3in

%
 \font\twelvebf=cmbx12
 \font\twelvett=cmtt12
 \font\twelveit=cmti12
 \font\twelvesl=cmsl12
 \font\twelverm=cmr12		\font\ninerm=cmr9
 \font\twelvei=cmmi12		\font\ninei=cmmi9
 \font\twelvesy=cmsy10 at 12pt	\font\ninesy=cmsy9
 \skewchar\twelvei='177		\skewchar\ninei='177
 \skewchar\seveni='177	 	\skewchar\fivei='177
 \skewchar\twelvesy='60		\skewchar\ninesy='60
 \skewchar\sevensy='60		\skewchar\fivesy='60
%
%

%
 \font\fourteenrm=cmr12 scaled 1200
 \font\seventeenrm=cmr12 scaled 1440
 \font\fourteenbf=cmbx12 scaled 1200
 \font\seventeenbf=cmbx12 scaled 1440
%
%

%
%
%
\font\tenmsb=msbm10
\font\twelvemsb=msbm10 scaled 1200
\newfam\msbfam

%
\font\tensc=cmcsc10
\font\twelvesc=cmcsc10 scaled 1200
\newfam\scfam

%
\def\seventeenpt{\def\rm{\fam0\seventeenrm}%
 \textfont\bffam=\seventeenbf	\def\bf{\fam\bffam\seventeenbf}}
\def\fourteenpt{\def\rm{\fam0\fourteenrm}%
 \textfont\bffam=\fourteenbf	\def\bf{\fam\bffam\fourteenbf}}
\def\twelvept{\def\rm{\fam0\twelverm}%
 \textfont0=\twelverm	\scriptfont0=\ninerm	\scriptscriptfont0=\sevenrm
 \textfont1=\twelvei	\scriptfont1=\ninei	\scriptscriptfont1=\seveni
 \textfont2=\twelvesy	\scriptfont2=\ninesy	\scriptscriptfont2=\sevensy
 \textfont3=\tenex	\scriptfont3=\tenex	\scriptscriptfont3=\tenex
 \textfont\itfam=\twelveit	\def\it{\fam\itfam\twelveit}%
 \textfont\slfam=\twelvesl	\def\sl{\fam\slfam\twelvesl}%
 \textfont\ttfam=\twelvett	\def\tt{\fam\ttfam\twelvett}%
 \scriptfont\bffam=\tenbf 	\scriptscriptfont\bffam=\sevenbf
 \textfont\bffam=\twelvebf	\def\bf{\fam\bffam\twelvebf}%
 \textfont\scfam=\twelvesc	\def\sc{\fam\scfam\twelvesc}%
 \textfont\msbfam=\twelvemsb	
 \baselineskip 14pt%
 \abovedisplayskip 7pt plus 3pt minus 1pt%
 \belowdisplayskip 7pt plus 3pt minus 1pt%
 \abovedisplayshortskip 0pt plus 3pt%
 \belowdisplayshortskip 4pt plus 3pt minus 1pt%
 \parskip 3pt plus 1.5pt
 \setbox\strutbox=\hbox{\vrule height 10pt depth 4pt width 0pt}}
\def\tenpt{\def\rm{\fam0\tenrm}%
 \textfont0=\tenrm	\scriptfont0=\sevenrm	\scriptscriptfont0=\fiverm
 \textfont1=\teni	\scriptfont1=\seveni	\scriptscriptfont1=\fivei
 \textfont2=\tensy	\scriptfont2=\sevensy	\scriptscriptfont2=\fivesy
 \textfont3=\tenex	\scriptfont3=\tenex	\scriptscriptfont3=\tenex
 \textfont\itfam=\tenit		\def\it{\fam\itfam\tenit}%
 \textfont\slfam=\tensl		\def\sl{\fam\slfam\tensl}%
 \textfont\ttfam=\tentt		\def\tt{\fam\ttfam\tentt}%
 \scriptfont\bffam=\sevenbf 	\scriptscriptfont\bffam=\fivebf
 \textfont\bffam=\tenbf		\def\bf{\fam\bffam\tenbf}%
 \textfont\scfam=\tensc		\def\sc{\fam\scfam\tensc}%
 \textfont\msbfam=\tenmsb	
 \baselineskip 12pt%
 \abovedisplayskip 6pt plus 3pt minus 1pt%
 \belowdisplayskip 6pt plus 3pt minus 1pt%
 \abovedisplayshortskip 0pt plus 3pt%
 \belowdisplayshortskip 4pt plus 3pt minus 1pt%
 \parskip 2pt plus 1pt
 \setbox\strutbox=\hbox{\vrule height 8.5pt depth 3.5pt width 0pt}}

%
\def\twelvepoint{%
 \def\small{\tenpt\rm}%
 \def\normal{\twelvept\rm}%
 \def\big{\fourteenpt\rm}%
 \def\huge{\seventeenpt\rm}%
 \footline{\hss\twelverm\folio\hss}
 \normal}
%

%
\def\bigbold{\big\bf}

%
\catcode`\@=11
%
%
\def\footnote#1{\edef\@sf{\spacefactor\the\spacefactor}#1\@sf
 \insert\footins\bgroup\small
 \interlinepenalty100	\let\par=\endgraf
 \leftskip=0pt		\rightskip=0pt
 \splittopskip=10pt plus 1pt minus 1pt	\floatingpenalty=20000
 \smallskip\item{#1}\bgroup\strut\aftergroup\@foot\let\next}
%
%
%
%
\def\hexnumber@#1{\ifcase#1 0\or 1\or 2\or 3\or 4\or 5\or 6\or 7\or 8\or
 9\or A\or B\or C\or D\or E\or F\fi}
\edef\msbfam@{\hexnumber@\msbfam}

%
%
%
\catcode`\@=12

\newcount\EQNO      \EQNO=0
\newcount\FIGNO     \FIGNO=0
\newcount\REFNO     \REFNO=0
\newcount\SECNO     \SECNO=0
\newcount\SUBSECNO  \SUBSECNO=0
\newcount\FOOTNO    \FOOTNO=0
\newbox\FIGBOX      \setbox\FIGBOX=\vbox{}
\newbox\REFBOX      \setbox\REFBOX=\vbox{}
\newbox\RefBoxOne   \setbox\RefBoxOne=\vbox{}

\expandafter\ifx\csname normal\endcsname\relax\def\normal{\null}\fi

\def\Eqno{\global\advance\EQNO by 1 \eqno(\the\EQNO)%
    \gdef\label##1{\xdef##1{\nobreak(\the\EQNO)}}}
\def\Fig#1{\global\advance\FIGNO by 1 Figure~\the\FIGNO%
    \global\setbox\FIGBOX=\vbox{\unvcopy\FIGBOX
      \narrower\smallskip\item{\bf Figure \the\FIGNO~~}#1}}
\def\Ref#1{\global\advance\REFNO by 1 \nobreak[\the\REFNO]%
    \global\setbox\REFBOX=\vbox{\unvcopy\REFBOX\normal
      \smallskip\item{\the\REFNO .~}#1}%
    \gdef\label##1{\xdef##1{\nobreak[\the\REFNO]}}}
\def\Section#1{\SUBSECNO=0\advance\SECNO by 1
    \bigskip\leftline{\bf \the\SECNO .\ #1}\nobreak}
\def\Subsection#1{\advance\SUBSECNO by 1
    \medskip\leftline{\bf \ifcase\SUBSECNO\or
    a\or b\or c\or d\or e\or f\or g\or h\or i\or j\or k\or l\or m\or n\fi
    )\ #1}\nobreak}
\def\Footnote#1{\global\advance\FOOTNO by 1
    \footnote{\nobreak$\>\!{}^{\the\FOOTNO}\>\!$}{#1}}
\def\SameFootnote{$\>\!{}^{\the\FOOTNO}\>\!$}

\def\References{\bigskip\centerline{\bf REFERENCES}
                \smallskip\copy\REFBOX}
\def\NewRefPage{\setbox\RefBoxOne=\vbox{\unvcopy\REFBOX}
		\setbox\REFBOX=\vbox{}
		\def\References{\bigskip\centerline{\bf REFERENCES}
                		\nobreak\smallskip\nobreak\copy\RefBoxOne
				\vfill\eject
				\smallskip\copy\REFBOX}
		\def\NewRefPage{}}




\font\tenbm=cmmib10
\font\ninei=cmmi9
\newfam\bmfam

\def\tenpointbmit{
\textfont\bmfam=\tenbm
\scriptfont\bmfam=\seveni
\scriptscriptfont\bmfam=\fivei
\def\bmit{\fam\bmfam\tenbm}
}

\tenpointbmit

\mathchardef\Gamma="7100
\mathchardef\Delta="7101
\mathchardef\Theta="7102
\mathchardef\Lambda="7103
\mathchardef\Xi="7104
\mathchardef\Pi="7105
\mathchardef\Sigma="7106
\mathchardef\Upsilon="7107
\mathchardef\Phi="7108
\mathchardef\Psi="7109
\mathchardef\Omega="710A
\mathchardef\alpha="710B
\mathchardef\beta="710C
\mathchardef\gamma="710D
\mathchardef\delta="710E
\mathchardef\epsilon="710F
\mathchardef\zeta="7110
\mathchardef\eta="7111
\mathchardef\theta="7112
\mathchardef\iota="7113
\mathchardef\kappa="7114
\mathchardef\lambda="7115
\mathchardef\mu="7116
\mathchardef\nu="7117
\mathchardef\xi="7118
\mathchardef\pi="7119
\mathchardef\rho="711A
\mathchardef\sigma="711B
\mathchardef\tau="711C
\mathchardef\upsilon="711D
\mathchardef\phi="711E
\mathchardef\cho="711F
\mathchardef\psi="7120
\mathchardef\omega="7121
\mathchardef\varepsilon="7122
\mathchardef\vartheta="7123
\mathchardef\varpi="7124
\mathchardef\varrho="7125
\mathchardef\varsigma="7126
\mathchardef\varphi="7127



%
%
\twelvepoint			
%
%



\font\tenbm=cmmib10
\font\ninei=cmmi9
\newfam\bmfam

\def\tenpointbmit{
\textfont\bmfam=\tenbm
\scriptfont\bmfam=\seveni
\scriptscriptfont\bmfam=\fivei
\def\bmit{\fam\bmfam\tenbm}
}

\tenpointbmit

\mathchardef\Gamma="7100
\mathchardef\Delta="7101
\mathchardef\Theta="7102
\mathchardef\Lambda="7103
\mathchardef\Xi="7104
\mathchardef\Pi="7105
\mathchardef\Sigma="7106
\mathchardef\Upsilon="7107
\mathchardef\Phi="7108
\mathchardef\Psi="7109
\mathchardef\Omega="710A
\mathchardef\alpha="710B
\mathchardef\beta="710C
\mathchardef\gamma="710D
\mathchardef\delta="710E
\mathchardef\epsilon="710F
\mathchardef\zeta="7110
\mathchardef\eta="7111
\mathchardef\theta="7112
\mathchardef\iota="7113
\mathchardef\kappa="7114
\mathchardef\lambda="7115
\mathchardef\mu="7116
\mathchardef\nu="7117
\mathchardef\xi="7118
\mathchardef\pi="7119
\mathchardef\rho="711A
\mathchardef\sigma="711B
\mathchardef\tau="711C
\mathchardef\upsilon="711D
\mathchardef\phi="711E
\mathchardef\cho="711F
\mathchardef\psi="7120
\mathchardef\omega="7121
\mathchardef\varepsilon="7122
\mathchardef\vartheta="7123
\mathchardef\varpi="7124
\mathchardef\varrho="7125
\mathchardef\varsigma="7126
\mathchardef\varphi="7127



\vskip 2cm


\centerline{\bigbold GRAVITY, GAUGES AND CLOCKS}\vskip 0.7cm

\bigskip\bigskip\bigskip

\centerline{Pierre Teyssandier }
\medskip
\centerline{\it Laboratoire de Gravitation et Cosmologie Relativistes,}
\centerline{\it  Universite Pierre et
Marie Curie, CNRS/URA769}
\centerline{\it Tour 22-12, BP142, 4 Pl. Jussieu, 75252 Paris,
Cedex 05}
\medskip
\medskip

\centerline{Robin W Tucker}
\medskip
\centerline{\it School of Physics and Chemistry,}
\centerline{\it University of Lancaster,
		 Lancs. LA1 4YB, UK}
\centerline{\tt r.tucker{\rm @}lancaster.ac.uk}

\vskip 1cm
\vskip 2cm

\bigskip\bigskip\bigskip\bigskip

\centerline{\bf ABSTRACT}
\vskip 1cm

\midinsert
\narrower\narrower\noindent


We discuss the definitions of standard clocks in theories of
gravitation. These definitions are motivated by the invariance of actions
under different gauge symmetries. We contrast the definition of a standard
Weyl clock with that of a clock in general relativity and argue that the
historical criticisms of theories based on non-metric compatible connections
by  Einstein, Pauli and others must be considered in the context of Weyl's
original gauge symmetry. We argue that standard Einsteinian clocks can be
defined in non-Riemannian theories of gravitation by  adopting the
Weyl group as a local gauge symmetry that {\it preserves the metric}
and discuss the hypothesis that atomic
clocks  may be adopted to measure proper time in the presence of
non-Riemannian gravitational fields. These ideas are illustrated in terms
of a recently developed model of gravitation based on a non-Riemannian
space-time geometry.

\endinsert




\vfill
\eject

\headline={\hss\rm -~\folio~- \hss}     

\def\frac#1#2{{#1\over #2}}
\def\frac#1#2{{#1\over #2}}

\Section{Introduction}

The strong and electroweak interactions between matter are described by
an action principle that encodes the strong and electroweak charges into
fields carrying representations of certain Lie groups. The fields are
 coupled
together in a manner
that renders an action functional invariant under local gauge
transformations.  To this end additional  Yang-Mills gauge connection fields
 are introduced as sections of principal
bundles over space-time.  By comparison
 Einstein's classical theory of gravitation is
formulated in the language of pseudo-Riemannian geometry.  The interaction of
gravitation and matter is encoded into a field theory in which matter is
described by tensor and spinor fields  and their derivatives
and the gravitational field is expressed in terms of a space-time metric and
its derivatives.  These fields are usually required to make some action
functional extremal.  Unlike the Yang-Mills action principle in which the
Yang-Mills gauge fields are dynamically independent, the connection used to
describe Einsteinian  gravity is constrained to be metric-compatible and
torsion-free.

The experimental evidence for Einstein's theory received
considerable enhancement with the recent observation of the rate of slowing
of the binary pulsar PSR 1913+16.
However it appears that certain other  astrophysical
observations do not rest so easily with classical gravitation.  In
particular some velocity distributions of stars in galaxies are hard to
reconcile with the observed matter distributions if they follow from
Newtonian dynamics.  Since Einstein's theory reproduces Newtonian gravity in
a non-relativistic weak-field limit this has led some to conjecture that
such galaxies may contain significant amounts of dark matter.  An
alternative explanation is that Newtonian dynamics requires modifications in
this context.  If this alternative is taken seriously it invites one to
consider  alternatives to Einstein's metric theory that may be testable in an
astrophysical domain.

Some of the earliest generalisations to Einstein's theory were entertained
by Cartan and Weyl.  The former suggested that the Levi-Civita connection
 used by Einstein  remained
metric-compatible but relaxed  to admit torsion while the latter made an
attempt to unify electromagnetism with gravity in terms of a theory based on
a non metric-compatible connection with zero torsion.  Although Weyl's
efforts proved abortive, modern string-inspired low energy effective actions
for gravity and matter can be formulated in terms of a non-Riemannian
connection with prescribed torsion and non-metricity,
\Ref{J Scherk, J H Schwarz, Phys. Letts {\bf 52B} (1974) 347},
\Ref{ T Dereli, R W Tucker, An Einstein-Hilbert Action for Axi-Dilaton
Gravity in \break 4-Dimensions, Lett. Class. Q. Grav. To appear},
\Ref{ T Dereli, M \"Onder,  R W Tucker, Solutions for Neutral  Axi-Dilaton
Gravity in 4-Dimensions, Lett. Class. Q. Grav. To appear},
\Ref{ T Dereli, R W Tucker,   Class. Q. Grav. {\bf 11} (1994) 2575}.
  It is difficult at
present to confront such effective theories with data since they are plagued
by an excess of unobservable scalar fields.

A less ambitious approach is to contemplate simple modifications of
Einstein's theory that do not a-priori constrain the geometry to
be pseudo-Riemannian  and seek actions with extrema
that dynamically prescribe the
metric and connection for the gravitational field.  There exists a large
literature that attempts to pursue this programme.  One of the earliest
efforts to ameliorate the difficulties that arise in the perturbative
quantisation of Einstein's theory of gravity contemplated a connection
constrained to be metric-compatible but with torsion expressed in terms of a
hypothetical gravitino field.  Theories based on actions in which the
connection is not constrained in the variational principle have been studied
but the actions have often suffered from being either ad-hoc or lacking in
motivation.  Furthermore, following Weyl's earlier attempts,  theories
with a non-metric compatible connection have often been viewed with
suspicion, partly as a result of the arguments made against Weyl's theory
by Einstein, Pauli and others
\Ref{W Pauli, ``Relativit\"atstheorie'', in Encyklop\"adie der
mathematischen Wissenschaften, Vol. 5, Physik, part 2 (B G  Teubner,
Leipzig, 1921)},
\Ref{ V P Vizgin, ``Einstein, Hilbert and Weyl: the Genesis of the
Geometric Unified Field Theory Program'', in Einstein Series, Vol 1:
Eimstein and the History of General Relatrivity, Ed. D Howard and J Stachel,
(Birkha\"user, 1989) 300}.
Since a number of authors are actively pursuing non-Riemannian prescriptions
\Ref{F W Hehl, J D McCrea, E W Mielke, Y Ne'eman: ``Metric-affine
gauge theory of gravity: field equations, Noether identities, world
spinors, and breaking of dilation invariance''.
Physics Reports, To appear (1995)
}\label\Hehl,
\Ref{ F W Hehl, E A Lord, L L Smalley, Gen. Rel. Grav. {\bf 13} (1981) 1037},
\Ref{P Baekler,  F W Hehl, E W Mielke  ``Non-Metricity and Torsion'' in
Proc. of 4th Marcel Grossman Meeting on General Relativity, Part A,
Ed. R Ruffini  (North Holland 1986) 277},
\Ref{V N Ponomariov, Y Obuchov, Gen. Rel. Grav. {\bf 14} (1982) 309},
\Ref{ J D McCrea, Class. Q. Grav. {\bf 9} (1992) 553},
\Ref{ A A Coley, Phys. Rev. {\bf D27} (1983) 728},
\Ref{ A A Coley, Phys. Rev. {\bf D28} (1983) 1829, 1844},
\Ref{ A A Coley, Nuov. Cim. {\bf 69B} (1982) 89},
\Ref{M Gasperini, Class. Quant. Grav. {\bf 5} (1988) 521},
\Ref{J Stelmach, Class. Quant. Grav. {\bf 8} (1991) 897},
\Ref{ A K Aringazin, A L Mikhailov, Class. Q. Grav. {\bf 8} (1991) 1685},
\Ref{ J-P Berthias, B Shahid-Saless, Class. Q. Grav. {\bf 10} (1993) 1039}
and there is the possibility of confronting these with the predictions of
Einstein's theory in the context of astrophysical data it is worthwhile
investigating the nature of these historic criticisms against non-Riemannian
descriptions of gravitation.  It is the purpose of this note to put the
arguments of Einstein and Pauli into perspective and argue that such
criticisms need to be used with caution when discussing theories based on
non-metric connections.  We exploit the language of modern differential
geometry since this offers the most precise way to express the basic notions
involved.

\def\RR{{\bf R}}
\def\gg{{\bf g}}
\def\SS{{\bf S}}
\def\TT{{\bf T}}
\def\AA{{\bf A}}
\def\FF{{\bf F}}

\def\nn{{\bf \nabla}}

\Section{Non-Riemannian geometry}

Let us briefly recall that a non-Riemannian space-time geometry
is defined by a pair  $(\gg,\nn)$    where  $\gg$
 is a metric tensor with Lorentzian
signature  and   $\nn$     is
a general linear  (Koszul)   connection.  From this pair  one can construct
$\SS=\nn \gg$      the
gradient of $\gg$, $\TT$, the torsion tensor      and $\RR_{X,Y}$,
 the curvature operator.
 If  $X,Y,Z$ are    arbitrary vector fields on space-time then
$\TT(X,Y)=\nn_XY -\nn_Y X -[X,Y]$ and $\RR_{X,Y} Z=\nn_X\nn_Y Z -\nn_Y\nn_X Z
-\nn_{[X,Y]}Z$.   If  $\nn$   is
chosen so that $\SS=0$ and $\TT=0$  the geometry is pseudo-Riemannian
and the gravitational
field is associated with the Riemann  curvature tensor $\RR$ where
$\RR(X,Y,Z,\beta)=\beta(\RR_{X,Y} Z)$ for an arbitrary 1-form $\beta$.
       This tensor is then
determined solely by the metric and the connection is called the Levi-Civita
connection.
Just as the   Levi-Civita  connection of a pseudo-Riemannian
geometry can be expressed
in terms of  $\gg$   alone a  general $\nn$
can be expressed in terms of $\gg, \TT$ and $\SS$:
$$2\gg(Z,\nn_{X}Y)=X(\gg(Y,Z))+Y(\gg(Z,X))-Z(\gg(X,Y))-$$
$$\quad\quad \gg(X,[Y,Z])-\gg(Y,[X,Z])-\gg(Z,[Y,X])-$$
$$\quad\quad\quad \gg(X,\TT(Y,Z))-\gg(Y,\TT(X,Z))-\gg(Z,\TT(Y,X))-$$
$$\quad\quad \SS(X,Y,Z)-\SS(Y,Z,X)+\SS(Z,X,Y)\Eqno $$\label\conn\
A geometry with non-zero torsion and a non-metric compatible
connection is also refered to as a ``metric-affine'' geometry by some
authors \Ref{A A Coley, Gen. Rel. Grav.
{\bf 14} (1982) 1107, {\bf 16 (1984) 459}}.

\Section{Weyl's Theory}

In 1918  Weyl
\Ref{ H Weyl, Sitzungsber. d. Kgl. Preuss. Akad. d. Wiss. Berlin, p.465
(1918). (English Translation in {\bf Lorentz, Einstein, Minkowski and Weyl,
 The Principle of Relativity}, Dover) },
\Ref{ H Weyl, Ann. d. Physik {\bf 59}, (1919) 101},
\Ref{ H Weyl, {\bf Space, Time, Matter} {Methuen}, (1922) },
\Ref{ H Weyl, Geometrie und Physik, Naturwissenschaften {\bf 19} (1931) 49}
proposed a theory  based on an action functional $S[\gg,\AA]$
  where $\nn$     was a non-Riemannian
connection constrained to have
$$\TT=0\Eqno $$
$$\SS= \AA \otimes \gg\Eqno $$
     for some  1-form $\AA$.
Since his action was
invariant under the substitutions:
$$\gg\to e^\lambda\gg\Eqno $$\label\gaugeA
$$\AA\to\AA +d\lambda\Eqno $$\label\gaugeB
for any 0-form $\lambda$,
this theory determined a class of solutions
 $[\gg, \AA]$;  elements being equivalent under what
Weyl termed the  {\it gauge transformations} \gaugeA,\ \gaugeB.\
Classical observables predicted by this theory should be gauge invariant.
In Einstein's pseudo-Riemannian description of gravitation a standard
clock  is
modeled by any time-like curve  $C$ parametrised
   with a tangent vector $\dot C$
   of constant length $\sqrt{(-\gg(\dot C,\dot C))}$.
 Such a clock can be calibrated to measure  {\it proper time}
$\tau$  with a standard
rate  independent of $C$, by fixing the parametrisation of $C$ so that
$$\gg(\dot C,\dot C)=-1\Eqno $$\label\time
      (in a metric with signature $(-,+,+,+)$).
The notion of a standard
clock makes precise the notion of a freely falling observer, namely an
affinely parametrised autoparallel (geodesic) integral curve of the
Levi-Civita connection. Since  such a connection is compatible with a
prescribed metric, ($\SS=0$), the normalisation of $\dot C$
is preserved for any $C$.
Thus although
the elapsed proper time between events connected by $C$ is path dependent, any
particular standard Einsteinian
clock admits a proper time parametrisation independent of its
world line.
In Weyl's geometry no particular $\gg$
  in the class $[\gg, \AA]$ is preferred so
the identification of a clock as a device for {\it measuring}
proper time requires more care.  The condition \time\
    is not invariant under Weyl's gauge group.  However in Weyl's geometry
$\TT=0$ and  it then follows from
\conn\    that under the transformations  \gaugeA,\ \gaugeB\
Weyl's connection $\nn$
  remains invariant. Thus if  $\gg$ and $\AA$ determine
 $\nn^{[\gg,\AA]}$
and $\gg=e^\lambda\gg_1,\,\, \AA=\AA_1+d\lambda$ then for any $X,Y,Z$:
$$\gg(Z,\nn^{[\gg_1,\AA_1]}_X Y)=\gg(Z,\nn^{[\gg,\AA]}_X
Y).\Eqno$$\label\invar

\def\class{$[\gg,\AA]$}

A definition of a standard clock should then refer to the gauge invariant
connection rather than the  gauge non-invariant condition \time\ .
Thus one may model a Weyl standard clock to be a time-like (with respect to
any $\gg$    in \class) curve $C$    such that
$$\gg(\nn_{\dot C}\dot C,\dot C)=0.\Eqno $$\label\Weyl\
This condition is manifestly
gauge invariant under \gaugeA\ and \gaugeB\ .
It follows that for each time-like curve
there exists a standard clock parametrisation of $C$
 that is unique up to the affine reparametrisation
$$\tau\mapsto a\tau + b\Eqno$$
with real constants  $a$ and $b$
\Ref{ V Perlick, Class. Quantum Grav. {\bf 8} (1991) 1369}.
However if
  $d\AA\neq 0$   one cannot choose $a=1$   for all such curves.
(If this were possible one could construct a Weyl parallel  normalised
tangent  vector on any closed curve. That this is impossible with $d\AA\neq
0$ follows by differentiating \time\ with  $\nn$.) Thus the
relative {\it rates} of two such standard clocks depend on their relative
histories in general.
(This effect should not be confused with the dependence of elapsed time
betwen events produced by the difference in paths linking such events.)

If one assumes that a
standard clock in Weyl's geometry \class\  corresponds to
an atom emitting light of a definite frequency then two identical atoms that
diverged from a unique space-time event and returned to any later event,
could
not have the same frequency at such an event if
$\int_\Sigma d\AA\neq 0$,   where
$\Sigma$ is any world sheet bounded by the world lines of the two atoms.
 Weyl  attempted to identify $\FF=d\AA$ with the Maxwell electromagnetic field
before the $U(1)$ nature of the coupling to charged fields was recognised.
Hence the spectra emitted by atoms in an ambient electromagnetic field
would be predicted to depend on their histories
contrary to observation.
This was the reason that his unified theory of
gravitation and electromagnetism fell prey to the early criticisms by
Einstein and Pauli.
Note however that such criticisms remain valid  whether or not
 $\FF$  is identified with the electromagnetic field.
They  rely only on the gauge invariant definition of the time
parametrisation of a Weyl standard clock and the assumed correspondence
of an atomic
spectral line with the rate associated with such a clock.

\Section{Einsteinian Clocks}

It is of interest to compare the above identification of a Weyl standard clock
with an atomic clock  with that used in
 Einstein-Maxwell relativity.  Since this theory works with a unique
classical pair $\{g,F\}$
 satisfying the Einstein-Maxwell field equations,
where $F$ is the Maxwell 2-form,
 one identifies
the angular frequency $\omega[C_e]$ of an emitted electromagnetic
wave from any timelike world-line  $C_e$,     with the scalar product
$$\omega[C_e]=-\gg(\dot C_e,\dot\Gamma_p)\vert_{C_e}\Eqno $$
where $\gg(\dot C_e,\dot C_e)=-1$ and $\dot\Gamma_p$
   is tangent to an integral curve of the null field
  $\Gamma$   that intersects $C$   at the event $p$:
$$\gg (\dot\Gamma_p,\dot\Gamma_p)=0.\Eqno $$\label\null
 If the vector field $\Gamma$   is identified
with  $\gg(d\phi,-)$ where $\phi$
is the phase of the emitted electromagnetic wave
it follows from the identity
$$(\nn_X d\phi)(Y)=(\nn_Y d\phi)(X)\Eqno $$
for any vector fields $X,Y$ and Levi-Civita $\nn$
   that $$\nn_{\dot\Gamma}{\dot\Gamma}=0\Eqno $$    i.e. the integral curves of
${\Gamma}$     are affinely
parametrised null geodesics.
It is a basic assumption in Einsteinian relativity
that the frequency
of an atomic clock does not depend on its world line and so one chooses a
parametrisation of $\Gamma$ that keeps the emitted frequency constant.
If the emitted electromagnetic wave is
detected by some observer world-line $C_o$ with $ \gg(\dot C_o,\dot C_o)=-1$
     the observed angular frequency is
$$\omega[C_o]=-\gg(\dot C_o,\dot\Gamma)\vert_{C_o}\Eqno $$
    which may not be constant along $C_o$.
It is traditional to measure the ratio
$$1+z=\frac{\gg(\dot C_e,\dot\Gamma_e)} {\gg(\dot C_o,\dot\Gamma_o)}\Eqno $$
       since
any affine parametrisation of the null
geodesic joining the emission event to the observation event can
then be employed.

Thus the criticism of Weyl's theory is essentially
based upon the notion used
for identifying standard clocks.  By contrast to
Einstein's theory which works with a well defined metric, the necessity of
making observables class invariant necessitates an alternative definition
of a standard clock.  The identification of atomic clocks with the
parametrisation of curves defined by \Weyl\
is responsible for the criticisms made by Einstein, Pauli and others.  Such
criticisms remain in force however one identifies the
metric-gradient field $\SS$  in Weyl's theory.
We   stress that the
essence of these criticisms lies in the fundamental
gauge symmetry associated with Weyl's action principle
 not with the identification of
$\FF$    with the Maxwell field.  This symmetry, in turn, follows
from Weyl's particular choice of a
non-Riemannian geometry having zero
torsion and
$\SS={\AA\otimes\gg}$.

\Section{Non-Riemannian Gravitational Fields}

Given the success of the gauge
description of the Yang-Mills interactions in which the connection
associated with any Yang-Mills gauge group is unconstrained in a variational
principle, a more natural approach to a non-Riemannian description of
gravitation is to seek a
purely gravitational action $S[\gg,\nn]$
that gives field equations determining a
unique metric  $\gg$ without constraining $\TT$    and $\SS$.
As with the Yang-Mills action principles that dynamically prescribe a class
of Yang-Mills connections
one may expect such an action
to determine a class of  Koszul  connections $[\nn]$
rather than a class of pairs $[\gg,\AA]$    as in
the Weyl's approach.

The 1-dimensional Weyl group offers perhaps the simplest gauge symmetry
for a non-Riemannian action principle. In this approach the action
$S[\gg,\nn]$ is invariant under
$$\gg\to \gg \Eqno $$\label\gnewgravity\
$$\nn\to \nn +df\otimes \Eqno $$\label\newgravity\
where $f$ is any 0-form and the action of $\nn$ on an arbitrary tensor is
induced from that given by the action on 1-forms in \newgravity.\
Note, by contrast to the gauge transformations \gaugeA,\ \gaugeB,\
that the metric tensor
remains inert under this transformation.
In a fixed local coframe the connection 1-forms $\Lambda^c{}_b$ of $\nn$
transform according to
$$\Lambda^c{}_b\to \Lambda^c{}_b + \delta^c{}_b\, df\Eqno $$
which may be recognised as Einstein's $\lambda$-transformation \Ref{ A
Einstein, in {\bf The Meaning of Relativity}, 6th edition, Appendix II,
Methuen, London, (1956)}.
It follows that the curvature scalar ${\cal R}$ associated with $\nn$ is
invariant under this gauge symmetry so the Einstein-Hilbert action remains
a candidate contribution to such an action.
However since the connection is
not constrained to be metric-compatible  the gauge-invariant 2-form defined by
$${\bf ric}(X,Y)=Trace(\RR_{X,Y}) .\Eqno $$
 is not necessarily zero. In the absence of matter a
 simple modification of  the Einstein-Hilbert action is
$$S[\gg,\nn]=\int (\kappa_1  \,\,{\cal R} \star 1 + \kappa_2\,\,
ric\wedge \star ric)\Eqno $$\label\action
 in terms of the Hodge map $\star$ of $\gg$
and coupling constants $\kappa_1, \,\kappa_2$.
The field equations that render this action extremal \Ref{R W Tucker, C
Wang,
Black Holes with Weyl Charge  and Non-Riemannian Waves, Class. Q. Grav.
(To Appear) 1995
} generate
a non-Riemannian geometry with
$$\SS=\frac{1}{4}\, \AA \otimes \gg,\Eqno $$\label\torsion
$$\TT=-\frac{1}{4}\,(\AA \wedge e^a)\otimes X_a\Eqno $$\label\nonmetricity
where   the Weyl 1-form $\AA$ satisfies
$$d\star d\AA=0\Eqno$$
and $\{e^a\}$ and  $\{X_b\}$ is a pair of locally dual bases.

\def\Ein{{\bf Ein}}
Unlike Weyl's theory however, the metric $\gg$ is uniquely determined in any
chart
by the Einstein equation
$$\kappa_1 \Ein=T^{ric}\Eqno $$\label\Eineqn
 with a stress tensor $T^{ric}$ of purely geometric origin:
$${T^{ric}}_{ab} = -\kappa_2(4\,\hbox{ric}_{ac}\,\,\hbox{ric}^c{}_b +
             g_{ab}\,\hbox{ric}_{cd}\,\,\hbox{ric}^{cd}).\Eqno $$
In fact it follows from \torsion\ and \nonmetricity\ that the Einstein tensor
$\Ein$ in \Eineqn\ coincides with the Einstein tensor constructed with the
Levi-Civita connection and that any traditional electro-vac solution to the
Einstein-Maxwell equations will also solve \Eineqn.\   Although the additional
gravitational fields $\TT$ and $\SS$ are not zero the whole theory can be
recast in terms of the metric-compatible torsion-free Levi-Civita
connection.
The gauge symmetry \newgravity,\  \gnewgravity\
 is compatible with the definition of a standard
 Einsteinian clock to define proper time.

\Section{Conclusion}

In general relativity, the
identification of  proper time with the time indicated by an atomic
clock is often regarded as due to the ``local position invariance'' that is
embodied in the Einstein equivalence principle \Ref{ C Will, {\bf Theory and
Experiment in Gravitational Physics}, paras 2.3, 2.4, Cambridge University
Press, (1981)}.  The mathematical expression of this principle is well
known: the laws of physics that include the effects of Einsteinian gravity
exploit only the metric of a  pseudo-Riemannian geometry.
Modulo interactions with curvature they can often be
 obtained from their special relativistic form by replacing the
Minkowski metric by a general pseudo-Riemannian metric and expressing
derivatives as covariant derivatives with respect to the associated Levi-Civita
connection.
However in a generalised theory of gravitation this principle
of ``universal coupling to the metric''  must be abandoned.
The nature of the gravitational coupling to matter is dictated solely by an
action principle.
As a consequence, some matter fields
may couple not only to the metric but also to torsion and the metric
gradient, violating the above ``local position invariance''.
In Weyl's original  action principle the gauge symmetry \gaugeA,\  \gaugeB\
 prohibits
the identification of the parametrisation \time\  with measured time along a
worldline since it is not a gauge invariant condition. A standard clock in
Weyl's theory can, however,
be constructed with the aid of the gauge invariant condition \Weyl\
  but such a clock cannot in general be identified with a unique
parametrisation for its  world line.
The criticism's leveled at Weyl's theory arose from the hypothesis that
standard Weyl clocks could be identified with atomic clocks.
In an  action principle based on invariance under the gauge transformations
 \gnewgravity,\ \newgravity\ it is the metric that remains
invariant rather than the
connection so there is nothing to prevent one from defining a standard
clock as in Einstein's theory. If the matter composing an atomic clock has no
significant interaction with the torsion or the metric gradient of the
non-Riemannian gravitational field then it would be reasonable to identify
such a clock with a standard Einsteinian clock.

In both Weyl's
theory and one based on the gauge symmetry \gnewgravity,\ \newgravity\
an appropriate gauge invariant definition of a standard clock can be made.
Only when this is done can the
predictions of any theory be compared with the results of measurements
made with physical clocks.
In  Weyl's theory the adoption of standard Einsteinian clocks is prohibited.
On the other hand
experiment suggests that an atomic clock
 offers a robust  device for the
measurement of time.  If space-time admits
non-Riemannian gravitational fields based on the gauge symmetry
\gnewgravity,\ \newgravity\
one may experimentally check whether  such a device
is insensitive to the
torsion and metric gradient fields of its {\it local environment} and
whether it may  consequently be
modelled by a standard Einsteinian clock.
The relevance of these additional gravitational fields
 to large scale astrophysical problems and their interaction with matter
in general  is
under current investigation.

\Section{ Acknowledgment}

RWT is grateful to C Wang and R Kerner for discussions and to
R Kerner for providing facilities at the
Laboratoire de Gravitation et Cosmologie Relativistes, Universite Pierre et
Marie Curie, CNRS,  Paris where this work was begun and
to the Human Capital and Mobility Programme of the European Union for
partial support.

\References

\bye

\bye